\newif\ifsubmode
\submodefalse

\newif\ifprintfig
\printfigtrue

\newif\ifemulate
\emulatetrue

\ifsubmode
  \documentstyle[12pt,aasms4,epsf]{article}
  \received{}
  \accepted{}
  \journalid{}{}
  \articleid{}{}
\else
   \ifemulate
     \documentstyle [emulateapj,epsf,graphicx]{article}
     \submitted{{\it Accepted for publication in AJ}}
   \else
     \documentstyle[11pt,aaspp4,epsf]{article} 
     \slugcomment{{\it Accepted for publication in AJ}}
   \fi
\fi

\lefthead{Willman et al}
\righthead{High-Velocity Clouds in the EDR}

\def\lesssim{\mathrel{\hbox{\rlap{\hbox{\lower4pt\hbox{$\sim$}}}\hbox{$<$}}}}
\def\gtrsim{\mathrel{\hbox{\rlap{\hbox{\lower4pt\hbox{$\sim$}}}\hbox{$>$}}}}

\begin{document}

\title{SDSS Survey for Resolved Milky Way Satellite Galaxies II: High-Velocity Clouds in the Early Data Release}

\author{Beth Willman\altaffilmark{\ref{Washington}}, Julianne Dalcanton\altaffilmark{\ref{Washington}}$^,$\altaffilmark{\ref{Sloan}}, \v{Z}eljko Ivezi\'{c}\altaffilmark{\ref{Princeton}}, Donald P. Schneider\altaffilmark{\ref{PSU}}, Donald G. York\altaffilmark{\ref{Chicago}}
}


\newcounter{address}

\setcounter{address}{1}
\altaffiltext{\theaddress}{University of Washington, Department of Astronomy,
Box 351580, Seattle, WA 98195
\label{Washington}}
 
\addtocounter{address}{1}
\altaffiltext{\theaddress}{Alfred P. Sloan Research Fellow
\label{Sloan}}

\addtocounter{address}{1}
\altaffiltext{\theaddress}{Princeton University Observatory, Princeton, NJ 08544
\label{Princeton}}

\addtocounter{address}{1}
\altaffiltext{\theaddress}{Department of Astronomy and Astrophysics, The Pennsylvania State University, University Park, PA 16802
\label{PSU}}

\addtocounter{address}{1}
\altaffiltext{\theaddress}{University of Chicago, Astronomy \& Astrophysics
Center, 5640 S. Ellis Ave., Chicago, IL 60637
\label{Chicago}}


\ifsubmode\else
  \ifemulate\else
     \clearpage
  \fi
\fi


\ifsubmode\else
  \ifemulate\else
     \baselineskip=14pt
  \fi
\fi

\begin{abstract}
In this paper, we limit the stellar content of 13 high-velocity clouds (HVCs; including 1 compact HVC)  using the detection limits of a new survey for resolved Milky Way satellite galaxies in  the Sloan Digital Sky Survey Early Data Release (EDR).  Our analysis is sensitive to stellar associations within the virial radius of the Milky Way that are up to 50$\times$ fainter than the faintest known Milky Way satellites. Statistically, we find no stellar overdensity associated with any of the clouds.  These non-detections suggest lower limits of M$_{HI}$/$L_V$ that range between $\sim 0.2$ and $250$, assuming cloud distances of 50 and 300 kpc and using fiducial optical scale lengths of 3$'$ and 7$'$.  We explore the implication of these non-detections on the origin of the HVCs.

\end{abstract}


\ifemulate\else
   \clearpage
\fi

\section{Introduction}
For decades, astronomers have struggled to determine the properties of the high-velocity clouds (HVCs), a population of objects identified by  21 cm HI emission  with velocities inconsistent with Galactic rotation.  Recent improvements in available radio survey data (e.g. the HI Parkes All-Sky Survey and the Leiden Dwingeloo Survey; Barnes et al.\ 2001 and Hartmann \& Burton 1997 respectively) and  UV instrumentation (e.g. STIS and FUSE), have substantially increased our understanding of these numerous ($\sim$ 2000 in the South alone, Putman et al.\ 2002) clouds.  However, the origin of the vast majority of the HVC population has remained controversial.  Several explanations for their presence have been suggested, including: a Galactic fountain; tidal or ram pressure stripping from dwarf galaxies; and low mass dark matter halos predicted by currently favored cosmologies but not observed in optical surveys (Blitz et al. 1999, Braun \& Burton 2000, among others).  A combination of these, and other, scenarios is likely necessary to explain the wide range of observed HVC properties (see Wakker \& van Woerden 1997 for overview of suggested origins and discussion).

One of the principle limitations in distinguishing between these scenarios is the unknown distances to the clouds (e.g. Gibson et al. 2001).   The most reliable distance estimates to date have resulted from absorption line studies toward background stars at known distances placing concrete upper and lower limits on the distances to a handful of nearby HVCs (Wakker 2001). Unfortunately, this method cannot be applied to clouds at extragalactic (i.e. $>$ 100 kpc) distances due to a lack of background stars. 

Alternatively, if the HVCs host a detectable stellar population, their distances could be measured directly  using the location of the tip of the Red Giant Branch or RR Lyrae stars. Identifying stellar populations associated with HVCs would therefore provide the measurements of both stellar content and distance necessary for constraining their possible origins. In particular, stellar associations would provide strong support for the theory that some HVCs may be extremely low mass galaxies and, as such, may likely host stars (see \S4.2 for discussion).

Several groups are currently conducting deep, pointed surveys toward compact HVCs (CHVCs) in an attempt to identify resolved stars associated with the clouds (Grebel, Braun \& Burton 2000, Gibson et al. 2001). Although these observations are excellent probes for stars close to the position of the cloud, there are notable drawbacks to this method. Observations of Local Group dwarf spheroidals (dSphs) suggest that HI associations with dSphs are not necessarily coincident with the galaxies' positions ( Carignan et al 1998, St-Germain et al. 1999, Blitz \& Robishaw 2000) . Furthermore, although only CHVCs have been targeted in existing searches, Putman \& Moore (2001) have shown that the CHVC distribution is not consistent with the predicted distribution of dark matter halos around the Milky Way and that other subsets of the HVC family possess properties more similar to those predicted for the halos.  

The only existing large-area search for stellar counterparts used digitized POSS plates to examine one square degree areas around 264 clouds (Simon \& Blitz 2002). Although their survey is sensitive to stellar associations throughout the Local Group, all 264 clouds were classified as non-detections.  However, the survey used a nonuniform data set with a poorly characterized surface brightness limit that is no fainter than the known Local Group dwarf galaxies. 

Substantial improvements in survey sensitivity and uniformity can be made using the Sloan Digital Sky Survey (SDSS). SDSS is a unique tool with which to search for stellar counterparts to HVCs, providing the means both to survey sufficiently large areas to detect displaced stars and to efficiently sample a large number of clouds. We are currently using SDSS data to conduct a survey for resolved low surface brightness satellites out to the Milky Way's virial radius (Willman et al. 2002, hereafter Paper I).  Our survey also provides a way to statistically investigate the possibility of stellar associations with the high-velocity clouds on very large scales.  Our detection limits (calculated in Paper I) provide a template with which to interpret the results, whether positive or null, in a quantitative framework.

In this paper, we present an analysis of 325 square degrees of the SDSS Early Data Release (EDR; Stoughton et al. 2002) imaging database. This zero-declination strip contains 13 zero-declination HVCs, as cataloged by Putman et al. (2002).  We briefly summarize our survey technique in \S\ref{sec:tech}, discuss the analysis of the SDSS Early Data Release in \S\ref{sec:application}, use the analysis results and survey detection limits to constrain the stellar nature of the HVCs in \S\ref{sec:constraints}, and discuss our findings in \S4.

\section{Our Survey for Galaxies Within $R_{virial}$}

\subsection{The SDSS Early Data Release}
The Sloan Digital Sky Survey (York et al. 2000) will eventually image nearly one-quarter of the sky.  To date, roughly 460 deg$^2$ of imaging data have been released to the general community in the Early Data Release (Stoughton et al. 2002). These imaging data are taken in drift-scan mode with a camera consisting of 30 2048x2048 CCDs (see Gunn et al. 1998) on a dedicated 2.5 m telescope at Apache Point Observatory.  The sky is observed nearly simultaneously through 5 broad-band filters ($u, g, r, i$ and $z$, Fukugita et al. 1996) and the data are $95\%$ complete for stellar sources to limiting magnitudes of roughly 22.0, 22.2, 22.2, 21.3 and 20.5 respectively (Stoughton et al. 2002). As the SDSS photometric system is not yet finalized, the EDR magnitudes used in this paper will be denoted as $u^*$,$g^*$,$r^*$,$i^*$, and $z^*$.  The current system will differ absolutely from the final system by only a few percent in every band except for $u$, where it will differ by no more than $10\%$ (Stoughton et al. 2002).  A combination of accurate astrometry (see Pier et al. 2002 for details on the astrometric calibration), color and morphological information produces robust star-galaxy separation to a limiting magnitude of r$^*$ = 21.5 (Ivezi\'{c} et al. 2000).  See Hogg et al. (2001) for details on the photometric calibration of the data and Smith et al. (2002) for a description of the photometric system.  Because we are looking for extragalactic objects, all magnitudes presented in this paper have been corrected for reddening due to Galactic extinction with the maps of Schlegel, Finkbeiner \& Davis (1998).

\subsection{Survey Technique} \label{sec:tech}
Our survey takes advantage of SDSS's major strengths (large sky coverage, uniformity, depth, and accurate photometry in five bands) to uniformly probe for very low surface brightness (LSB) dwarf galaxies, based only on resolved stars.  Paper I describes both our survey technique and detection limit calculations in detail, but we briefly summarize our method here.

SDSS's faint photometric limits and robust star-galaxy separation are such that about a magnitude of a stellar population's red giant branch can be resolved out to a distance of 350 kpc, roughly equivalent to the Milky Way's virial radius.  This allows us to identify candidate dwarf galaxies as spatial overdensities of resolved stars.  We apply a red color cut to stellar sources brighter than r$^*$ = 21.5.  The color cut, in g$^*$ - r$^*$ and r$^*$ - i$^*$, is designed to select stars whose colors are consistent with red giant branch colors.  Although our cut does not uniquely distinguish red giant branch stars from foreground red main sequence stars, it does reduce the stellar foreground by $\sim 75\%$, increasing the contrast of extragalactic sources. We then consider bins in r$^*$ magnitude, further reducing the foreground. Stars which pass the magnitude and color criteria are binned into a spatial density array, which is then smoothed with a circularly symmetric exponential filter with a scale length of either h=3$'$ or 7$'$.  Candidates are automatically identified as 5.5 and 5$\sigma$  statistical overdensities (for the 3$'$ and 7$'$ filter, respectively), in the smoothed sky map.  Each field of sky is analyzed multiple times, once for each $r^*$ magnitude bin.

In Paper I, we calculated our detection limits for purely old stellar populations with a range of angular scale lengths ($1' - 13'$) and distances (23 - 350 kpc).  Figure 5 in Paper I displays our survey's detailed detection limits for the stellar foreground at $\ell \sim 96^{\circ}, b \sim -60^{\circ}$.  The surface brightness limits  are a function of galaxy size and distance, and range between $ \mu_{V,0} \sim 26.7$ and $30.1$ mag arcsec$^{-1}$. (For comparison, Sextans is the lowest surface brightness Milky Way satellite, with $\mu_{0,V} = 26.2$ (Mateo 1998).)  Our detection efficiency rapidly drops off for small angular size ($< 2'$) systems, and is insensitive to systems more distant than $\sim$ 350 kpc.  The limits are not sensitive to seeing, because our technique uses only resolved stars brighter than the limit for robust star-galaxy separation.

\subsection{Application of the Survey to the EDR} \label{sec:application}

We have analyzed $\sim$ 325 square degrees of data in the SDSS EDR around zero declination between $350^{\circ} < \alpha_{2000} <50^{\circ}$ and $147^{\circ} < \alpha_{2000} <236^{\circ}$.  This region of sky contains 10 HVCs, 2 :HVCs (a cloud that could not unambiguously be identified as a CHVC),  and 1 CHVC as cataloged by Putman et al. (2002).  We have chosen the Putman et al. catalog as a reference for the present analysis because it is the most uniform and high resolution HVC catalog currently available.   Braun and Burton (1999) have also cataloged 1 CHVC in the zero-declination region covered by the EDR. The Braun and Burton CHVC is not coincident with the Putman CHVC, but rather is coincident with a Putman ``:HVC'' included in this analysis.
 
Our analysis produced 20 detections that we are currently investigating.  It is difficult to say conclusively that these stellar detections are not associated with any particular HVC, because the gas and stars in low mass Local Group galaxies are not necessarily spatially coincident.  However, we can show that the spatial displacements between the stellar detections and the HVCs are no different than we would expect for completely uncorrelated populations.  To test this, in Figure 1 we show a comparison of the distances between the detections and their nearest HVC centers with the distances between randomly generated positions and their nearest HVC centers.  A Kolmogorov-Smirnov test of these data shows that, statistically, none of the candidate detections are apt to be associated with any of the HVCs in the region.  This suggests that any stellar population associated with the HVCs must have a surface brightness fainter than the limits of our survey.

\placefigure{fig:disthist}

\section{Constraints on the Possible Stellar Content of the HVCs} \label{sec:constraints}

We now couple our null detections with the survey's detection limits to place quantitative limits on the stellar content of the HVCs, assuming they are located within our virial radius. As the detection limits vary with galaxy angular size and distance, we have selected two fiducial distances (50 and 300 kpc) and sizes (exponential scale lengths of 3' and 7$'$) for which to calculate the minimum HI mass-to-light ratios of the clouds.  We did not simply use the semi-major axes of the HVCs' 21 cm emission for the fiducial sizes because the radial extents of neutral hydrogen and stars within dwarf galaxies are often quite different (Young \& Lo 1997, Salpeter \& Hoffman 1996).

Table 1 displays both the HI and the constrained stellar properties of the 13 zero-declination HVCs covered in our analysis. Columns 1-5 of Table 1 contain the name, $\alpha_{2000}$ (in degrees), $\delta_{2000}$ (in degrees), peak $N_{HI}$ (in units of $10^{20} cm^{-2}$) and semi-major axis (in degrees) for each cloud as published in Putman et al. (2002).  Column 6 contains the assumed distance used to calculate: the semi-major axis in kpc (column 7); $M_{HI}$ (in units of $10^5 M_{\odot}$; column 8); and the optical detection limits for 3$'$ and 7$'$ sized systems (quantified as V-band central surface brightnesses of the old stellar population; columns 9 an 11).  $M_{HI}$ was  calculated using $M(HI) = .236\times D^2_{kpc}S$ (where S is the integrated line flux in Jy km sec$^{-1}$ from Putman et al.\ (2002)).  We have recalculated the detection limits from Paper I for a range of stellar foreground levels corresponding to the range of Galactic positions for the zero-declination HVCs.  

To get a sense of how the amount of HI contained in these HVCs compares to that contained in known gas-rich dwarf galaxies,  Figure 2 displays the HI masses of the zero-dec HVCs (for assumed distances of 50 and 300 kpc), the estimated HI masses of 9 gas-rich Local Group dSphs (Tucana -- Oosterloo et al. 1996; Phoenix -- St-Germain et al 1999; DDO 210, Leo I, LGS 3, Pegasus, Sextans, Sculptor -- Blitz \& Robishaw 2000; Antlia -- Barnes \& de Blok 2001), and the HI masses of the PSS-II LSB dwarf irregular (dIrr) sample of Schombert, McGaugh and Eder (2001).  We would like to note that the published HI detections of a few of the Local Group dSphs are not definitive associations, so the plotted HI masses are upper limits.  The 9 dSph HI detections in Figure 2 are either those that are conclusive associations or those that do not have conclusive evidence that shows otherwise.

From Figure 2, we see that the dIrr galaxies have much higher HI masses than the HVCs, and thus are not close analogs for a Local Group population of HVCs.  In contrast, the HVCs could contain HI masses that correspond to the approximate HI masses of the gas-rich Local Group dSphs.

\placefigure{fig:chvchistmass}

We calculate lower limits on the neutral hydrogen mass-to-light ratio ($M_{HI}$/$L_V$) of the HVCs (columns 10 and 12) using the limiting surface brightnesses from columns 9 and 11.  Specifically, we calculate a lower limit on the $M_{HI}/L_V$ for all of these clouds with $(.236\times S)/(10^6\times 100^{(-M_V/5 - 1 + M_{V,\odot}/5)})$, where we derive $M_V$ by integrating over an exponential galaxy profile normalized to the limiting central surface brightness.  M$_{HI}$ and luminosity both scale as $D^2$, so the limiting $M_{HI}$/$L_V$ ratio is independent of cloud distance, aside from the dependence of $\mu_{0,limit}$ on distance.

Figure 3 displays a comparison between our lower limits to M$_{HI}$/$L_V$ for zero-dec HVCs, the estimated M$_{HI}$/$L_V$  of Local Group dSphs and the M$_{HI}$/$L_V$  of the PSS-II dIrrs.  The available PSS-II data is in the I-band, but we convert to the corresponding V-band measurement for an old stellar population by using $<V-I> \sim 0.75$ for elliptical galaxies from Pildis et al. (1997).  

\placefigure{fig:chvchist}

The M$_{HI}$/$L_V$ lower limits for the HVCs range from about 0.2 to 250.  The calculated minimum $M_{HI}/L_V$ for many of the HVCs are similar to those of known gas-rich dwarf galaxies,  despite our faint detection limits (i.e. they may not be particularly faint in the optical compared to their very low HI masses).  Even so, Figure 3 shows that $\sim 40\%$ of the HVCs with assumed distances of 300 kpc have $M_{HI}/L_V$ greater than that of all known Local Group dSphs, and that 3/5 of that 40$\%$ also have $M_{HI}/L_V$ greater than the entire PSS-II dwarf sample.  If shown to be within dark matter halos,  these HVCs would be among the most gas rich self-gravitating galaxies observed to date.

\section{Implications of Non-Detections}

The import of a non-detection clearly depends on whether or not a cloud is closer than 350 kpc.  If  an HVC is more distant than 350 kpc, then we can place absolutely no constraints on its stellar content.  It is possible that a fraction of the HVCs in our sample are indeed at large distances.  Although upper distance limits from absorption line studies and observed head-tail morphologies would imply inner halo locations, we do not know of any such published analyses applied to any cloud in this sample (see Wakker 2001 for compendium of absorption line estimates; Br{\" u}ns et al.\ 2000; Quilis \& Moore 2001 ).


However, if one believes the many indirect estimates of HVC distances that place a large fraction of the clouds  closer than several hundred kiloparsecs (Bland-Hawthorn \& Maloney 1999, Weiner et al.\ 2001, Blitz et al.\ 1999, Quilis \& Moore 2001), it is logical to expect at least some of these clouds to be within our virial radius.

As further evidence that at least some of the 13 clouds are nearby, surveys for HI clouds outside of the Local Group have failed to identify starless clouds with M $\gtrsim 7 \times 10^6 M_{\odot}$ (Zwaan 2001, Zwaan \& Briggs 2000).  This observation suggests that the HVCs must either be sufficiently close that their HI masses fall below the Zwaan limit, or should host a stellar population detectable by existing Local Group surveys.  If located beyond our virial radius, four of the 13 HVCs have masses that rival the Zwaan limit but do not appear to have a detectable stellar population.  Therefore, it is quite likely at least all four of those clouds fall within our survey volume.

\subsection{Stars as a Distance Determination Tool for HVCs}

Clouds in our analysis closer than 350 kpc either have: {\it i)} no associated stars; {\it ii)} a stellar population fainter than our detection limits and a mass-to-light ratio, M$_{HI}$/L$_V$, higher than that given in Table 1; or {\it iii)} a stellar population with a very small angular scale length ($\lesssim 1'$)  but brighter than our given detection limits.  For clouds that do contain stars, as in cases {\it ii)} and {\it iii)}, most other survey methods would also not be able to efficiently uncover such faint or small populations. Pointed searches can probe deeper and to greater distances than large-scale surveys of resolved stars, but they are much less efficient and may have difficulty associating very small scale length systems with an HVC.  Large-scale surveys based of resolved stars could reach fainter limits only with prohibitively deeper exposures, with a priori assumptions about the undetected stellar populations (Odenkirchen et al.\ 2001), or with surveys for very specific stellar types (e.g. Horizontal Branch and RR Lyrae; Yanny et al.\ 2000, Ivezi{\' c} et al.\ 2000) that are only effective in the inner halo. Therefore, although our sample size is small, our detection limits are faint enough to conclude that stellar populations are unlikely to be an efficient or effective distance determination technique to HVCs within our virial radius.  A large-scale survey for overdensities of diffuse light using more recent and uniform data (such as SDSS) than available for past surveys (Simon \& Blitz 2002; Armandroff, Jacoby \& Davies 1998; Dalcanton et al.\ 1997) holds promise for efficiently improving existing limits on the possible stellar content of HVCs located beyond our virial radius.

\subsection{HVCs as ``Missing'' Galaxies}
There is a substantial body of circumstantial evidence to support the theory that a subset of high-velocity clouds inhabit some of the numerous low mass dark matter halos predicted by the current paradigm of structure formation, yet notably missing from our observations of the Local Group of galaxies (Blitz et al. 1999, Braun \& Burton 2000, among others).  For example, Blitz and Robishaw (2002) show that the masses and HI morphologies of many HVCs are consistent with the HI associated with Local Group dSphs. Furthermore, low mass analogues to known extremely gas-rich LSB galaxies (McMahon et al. 1990; O'Neil, Bothun \& Schombert 2000; Schombert, McGaugh \& Eder 2001) may easily have gone undetected in past surveys for Local Group galaxies.  

We can use existing  evidence to evaluate the likelihood of a gas-rich dark matter halo occurring within our survey's volume, and discuss the prospects for such halos to host stars. This will allow us to show that our non-detection of stars places few limits on missing galaxy explanations for the HVCs.

\subsubsection{Predictions From $N$-body Simulations}
$\Lambda$CDM simulations predict $\sim$ 300  sub-halos within the virial radius of the Milky Way with masses greater than or equal to the least massive known Milky Way companions ($v_c \geq 10$ km sec$^{-1}$, where $v_c \equiv (GM/r)^{1/2}$; Moore et al. 1999).  The expected number of halos increases steeply with decreasing mass, so $\sim$ 200 more sub-halos are expected with 5 km sec$^{-1} \lesssim v_c \lesssim 10$ km sec$^{-1}$  (Moore et al.\ 1999), and perhaps yet more at smaller masses.  Assuming a uniform distribution within the 325 square degree survey area, this translates to roughly 2-3 halos similar to those of known dwarf satellites, and a comparable number of lower mass halos.  Thus, it is reasonable to expect a small number of low mass dark matter sub-halos within our survey volume in a $\Lambda$CDM cosmology.  

These low mass, dark matter halos may or may not contain a detectable amount of neutral hydrogen. Numerous physical arguments can be invoked to suggest that v$_c \lesssim 30$ km sec$^{-1}$ halos (particularly those that contain stars) are unlikely to host neutral gas (Quinn, Katz \& Efstathiou 1996; Bullock, Kravtsov, \& Weinberg 2000; Ferrara \& Tolstoy 2000). However, empirical evidence suggests that low mass halos can host both stars and gas.  3 of the Milky Way's 9 dSphs (Leo I, Sculptor, and Sextans) potentially have HI associations (Blitz and Robishaw 2000).  All 3 of these dSphs are less massive than $v_c \sim 15$ km sec$^{-1}$ and closer than $\sim$ 250 kpc. The Sculptor detection is somewhat ambiguous because it is part of a larger cloud complex that may be mixed in with the Magellanic Stream (Putman 2000), and the Sextans detection is only a few sigma and has yet to be confirmed by other radio observations.  These existing observations imply that $\sim 10-30\%$ of Milky Way sub-halos are gas-rich.  Of course, the true gas-rich fraction is complicated by several observational effects, including: halos that are gas-rich due to inefficient star formation may not have been detected in Local Group surveys, and the lowest mass halos will more easily be stripped of their gas by feedback and tidal forces. Despite these uncertainties, it is reasonable to expect no more than 2-3 of the 4-6 sub-halos expected in our survey volume to be detectable as HVCs, if any (assuming Poisson statistics).  Thus, although it is possible that a dark matter bound cloud exists in our survey, the volume sampled is not yet large enough to make this a statistical certainty.

\subsubsection{Possible Stellar Content of Sub-Halos}
 
As discussed in \S4, any HI rich dark matter halo within our survey volume is either starless, contains a stellar population many times fainter than all known Milky Way satellites, or contains a very small scale length stellar population.  According to Benson et al. (2002), an M$_V = -7$ ($\sim 5\times$ fainter than the faintest known Milky Way dSph) galaxy is expected to have a half-mass angular radius of 1.5' at a distance of 300 kpc.  Therefore, the prospect of a small angular scale length system much brighter than our detection limits seems unlikely. With our small sample, it is neither productive nor possible to further consider which of those three options would be the most likely for any of these 13 clouds, were they shown to inhabit a dark matter halo. Studies which directly probe for dark matter associated with the clouds, such as that suggested by Lewis et al. (2000), may ultimately be the best test of theories that place HVCs within dark matter halos, although our current sample is too small to say one way or the other. 

It is quite possible that one/some of the predicted 2-6 gas-poor dark matter halos in the surveyed region do contain stars.  We are currently investigating whether any of our 20 detections actually represent a new low luminosity dwarf galaxy.

\section{Conclusion}
We have used our non-detections of stars associated with 13 HVCs to place new constraints on their stellar content.  We have shown that many of the 13 clouds have HI masses and (minimum) HI mass-to-light ratios similar to those of known dSphs, if the clouds are closer than 350 kpc.  Also, 3 of the HVCs have minimum M$_{HI}$/L that would place them among the most gas-rich galaxies, if shown to reside in a dark matter halo.  Given our results, and those of Simon and Blitz (2002), stellar populations appear unlikely to be a promising distance determination technique for a large number of HVCs.  However, the continued pursuit of stellar associations with HVCs is a worthwhile endeavor, as strong circumstantial evidence suggests that some fraction, however small, of the HVCs are likely to host a detectable low surface brightness (LSB) galaxy.  

As there are only 1-3 known gas-rich Milky Way dSph satellites, the discovery of {\it any} nearby gas-rich stellar association would substantially increase the sample and provide insights into extremely low-mass galaxy evolution.  We are currently pursuing the prospect that some of our 20 detections may actually be a new LSB HI poor Milky Way satellite. 

As our present survey volume is too small to unquestionably contain any of the ``missing'' low mass, dark matter halos predicted by $\Lambda$CDM cosmology, we cannot place limits on missing galaxy explanations for the HVCs. However, the analysis performed in this work will be applied to survey quality SDSS data as it becomes available, and then correlated with the current and future HVC catalogs of Putman et al. (HIPASS) and de Heij et al. (2002) (Leiden/Dwingeloo HI Survey). The completed Sloan Survey will cover $\sim 125\times$ more area than we have presented in this paper, guaranteeing that ``missing'' dark matter halos would be well represented in our survey volume (provided $\Lambda$CDM is correct).  Our future analysis of the SDSS will enable us to apply the arguments presented here to hundreds of clouds, a sample large enough for us to do a concrete statistical assessment of their nature.  


\acknowledgements
We would like to thank James Schombert for helping us access his dIrr catalogs, Mary Putman for pointing us toward her HVC catalog, and Christian Br{\" u}ns for sending us a list of the HVCs he observed to have head-tail morphologies.  

Funding for the distribution and creation of the SDSS Archive has been provided by the Alfred P. Sloan Foundation, the Participating Institutions, the National Aeronautics and Space Administration, the National Science Foundation, the U.S. Department of Energy, the Japanese Monbukagakusho, and the Max Planck Society. The SDSS Web site is http://www.sdss.org/.

The SDSS is managed by the Astrophysical Research Consortium (ARC) for the Participating Institutions.  The Participating Institutions are The University of Chicago, Fermilab, the Institute for Advanced Study, the Japan Participation Group, The Johns Hopkins University, Los Alamos National Laboratory, the Max-Planck-Institute for Astronomy (MPIA), the Max-Planck-Institute for Astrophysics (MPA), New Mexico State University, Princeton University, the United States Naval Observatory, and the University of Washington.


\ifsubmode\else
\baselineskip=10pt
\fi



\clearpage


\ifsubmode\else
\baselineskip=14pt
\fi


\newcommand{\figcapdisthist}{Cumulative distribution functions of the distances between our detections and the nearest HVC center and the distances between a random distribution of points and the nearest HVC center.\label{fig:disthist}}

\newcommand{\figcapchvchistmass}{Comparison between the HI masses for the HVCs with assumed distances of 50 and 300 kpc, the gas-rich Local Group dSphs (Oosterloo et al. 1996; St-Germain et al 1999; Blitz \& Robishaw 2000; Barnes \& de Blok 2001), and the PSS-II sample of dIrrs (Schombert, McGaugh and Eder 2001).  All of the dSphs and dIrrs in panels 2 and 3 would have easily been detected in our survey, were they located within the Milky Way's virial radius.\label{fig:chvchistmass}}

\newcommand{\figcapchvchist}{Comparison between the M$_{HI}$/L lower limits for the zero-declination HVCs from our analysis and the M$_{HI}$/L distribution of the  Local Group dSphs and the PSS-II sample of dIrrs (Schombert, Bothun and Eder 2001).\label{fig:chvchist}}


\ifsubmode
\figcaption{\figcapdisthist}
\clearpage
\else\printfigtrue\fi

\ifprintfig
\clearpage
\begin{figure}
\epsfxsize=14.0truecm
\centerline{\epsfbox{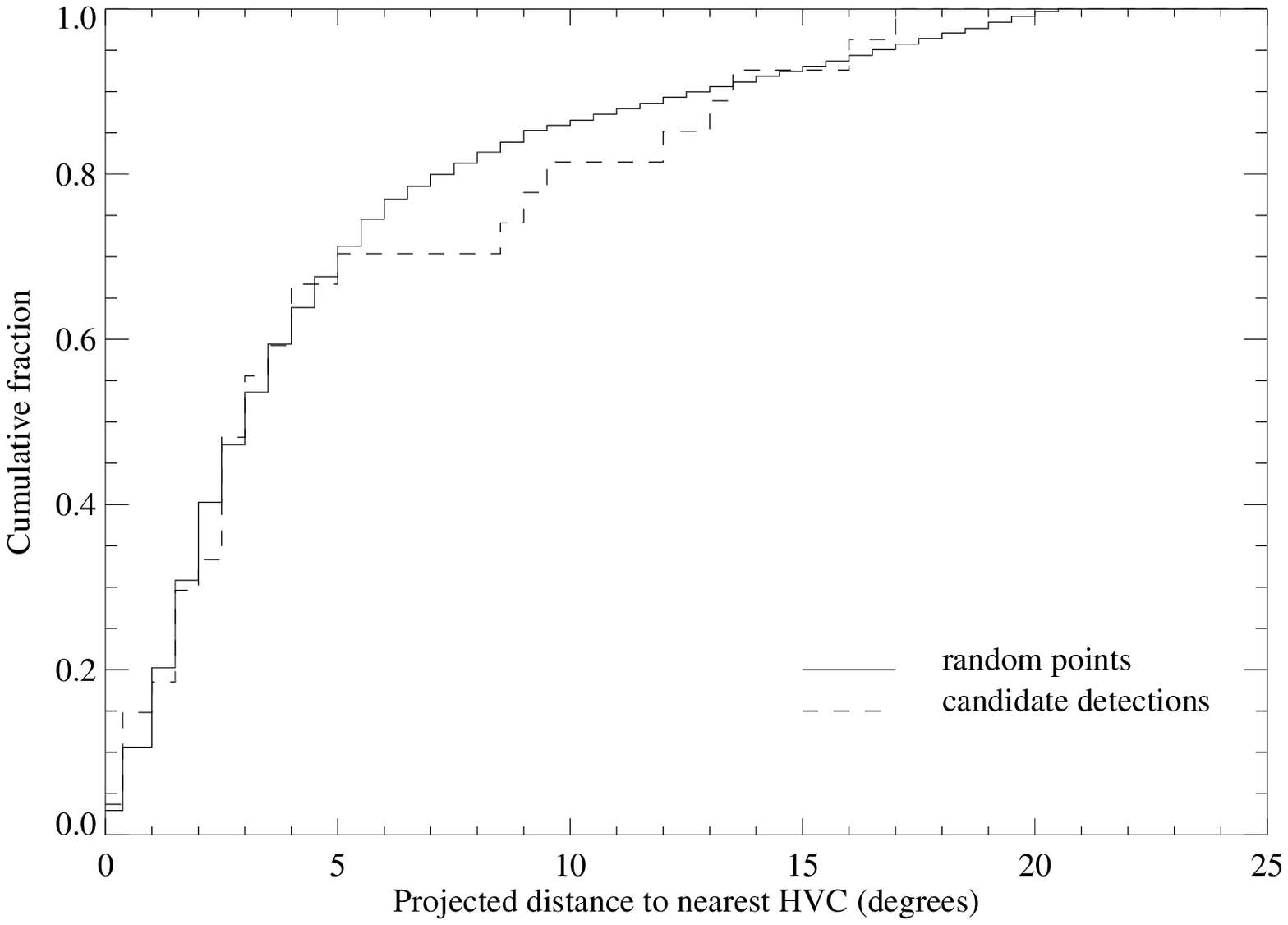}}
\ifsubmode
\vskip3.0truecm
\addtocounter{figure}{1}
\centerline{Figure~\thefigure}
\else\figcaption{\figcapdisthist}\fi
\end{figure}

\ifsubmode
\figcaption{\figcapchvchistmass}
\clearpage
\else\printfigtrue\fi

\ifprintfig
\clearpage
\begin{figure}
\epsfxsize=14.0truecm
\centerline{\epsfbox{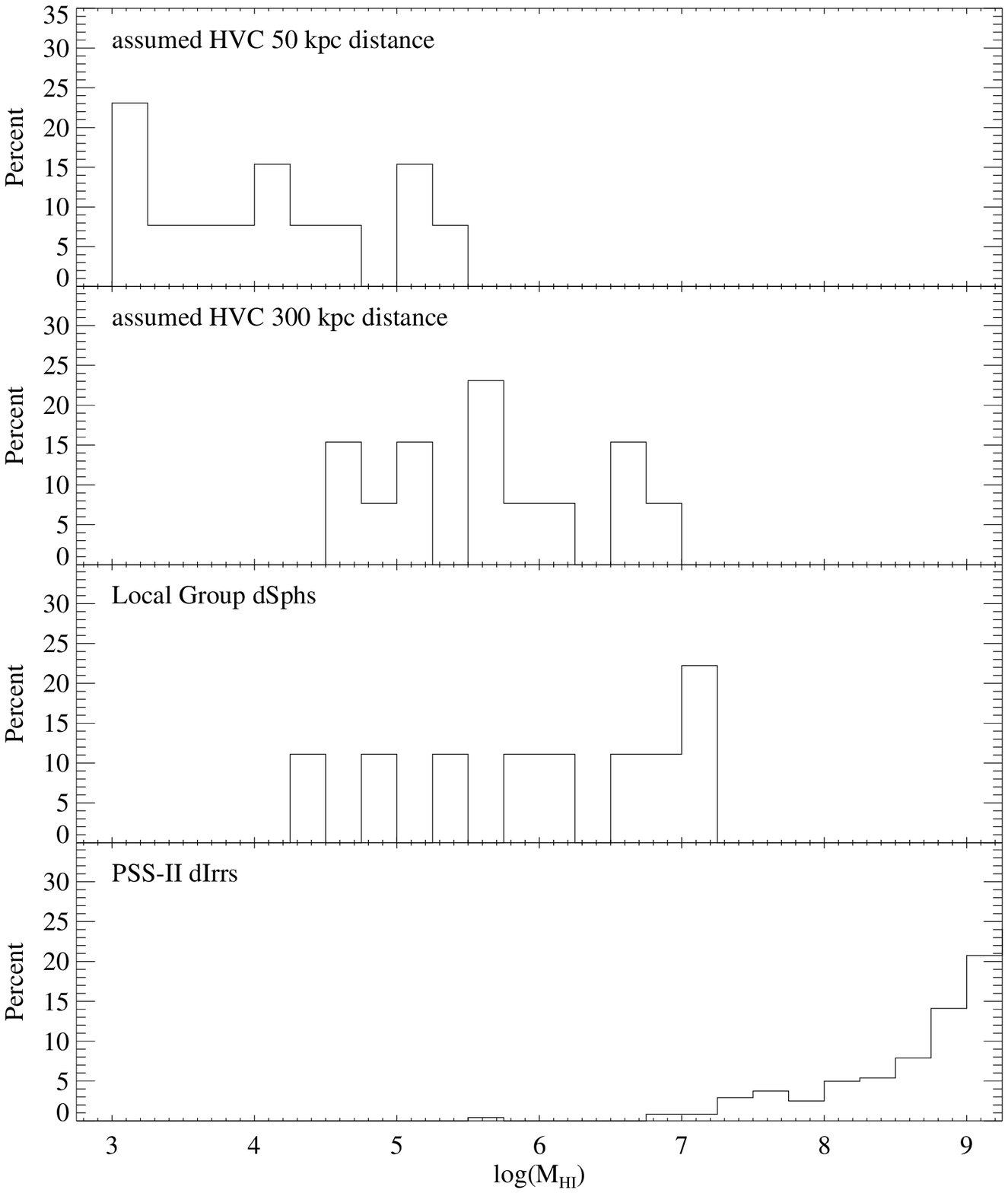}}
\ifsubmode
\vskip3.0truecm
\addtocounter{figure}{1}
\centerline{Figure~\thefigure}
\else\figcaption{\figcapchvchistmass}\fi
\end{figure}%
\clearpage

\ifsubmode
\figcaption{\figcapchvchist}
\clearpage
\else\printfigtrue\fi

\ifprintfig
\clearpage
\begin{figure}
\epsfxsize=14.0truecm
\centerline{\epsfbox{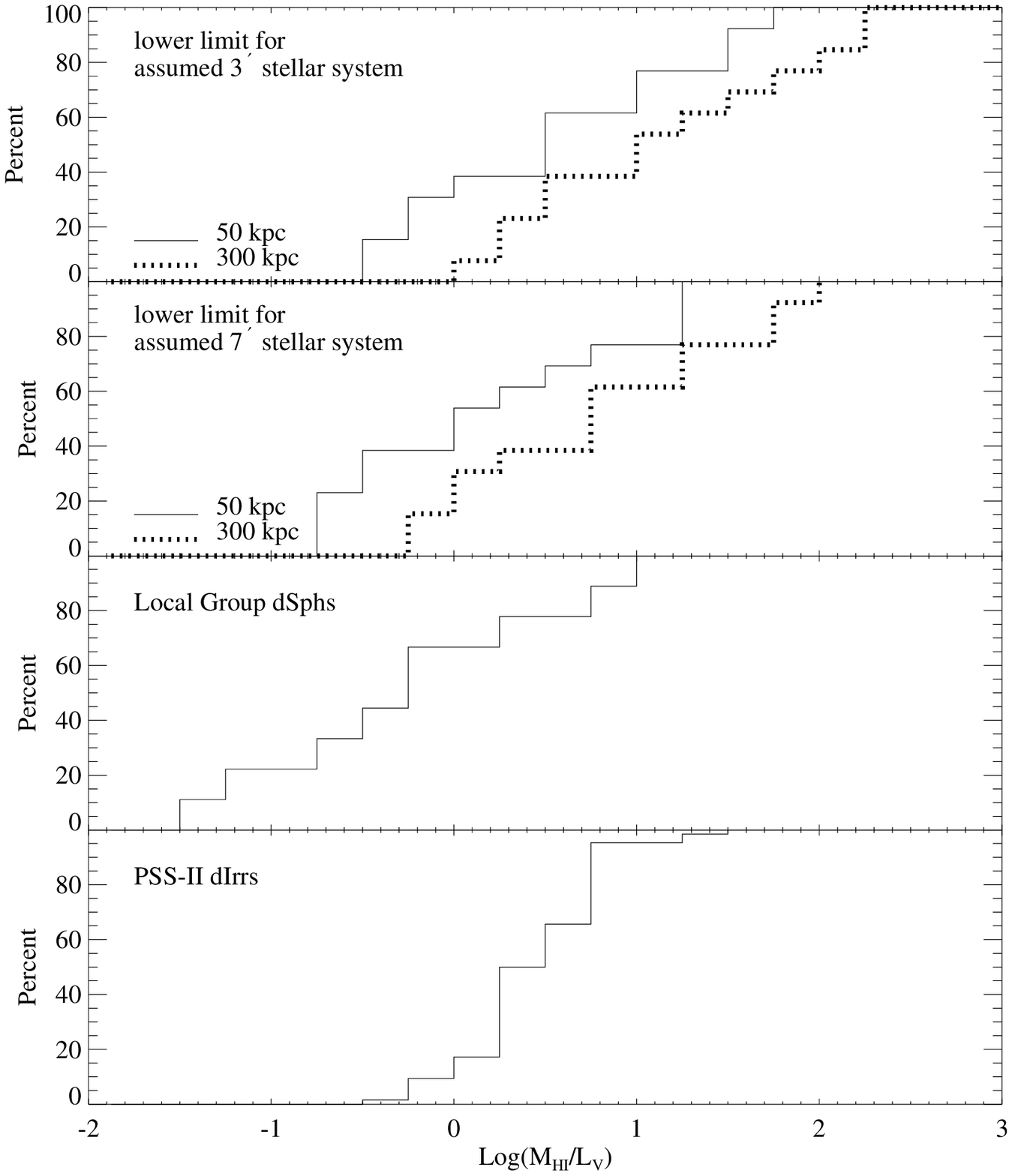}}
\ifsubmode
\vskip3.0truecm
\addtocounter{figure}{1}
\centerline{Figure~\thefigure}
\else\figcaption{\figcapchvchist}\fi
\end{figure}%
\clearpage




\clearpage

\scriptsize
\begin{deluxetable}{lrrrrrrrrrrr}
\tablewidth{0pt}
\tablecaption{Constraints on Zero-Declination HVC Properties}
\tablehead{
\colhead{Name\tablenotemark{a}} &  \colhead{$\alpha_{2000}$\tablenotemark{a}} & \colhead{$\delta_{2000}$\tablenotemark{a}} &
\colhead{$N_{HI}$\tablenotemark{a}}  & \colhead{MAJ\tablenotemark{a,b}} & \colhead{dist\tablenotemark{c}}  & \colhead{MAJ} & \colhead{$M_{HI}$} & \colhead{$\mu_{V,lim}$\tablenotemark{d}}  & \colhead{$\frac{M_{HI}}{L_V}$\tablenotemark{d}} & \colhead{$\mu_{V,lim}$\tablenotemark{d}} & \colhead{$\frac{M_{HI}}{L_V}$\tablenotemark{d} }  \\
\colhead{$ll\pm bb\pm vvv$}  & \colhead{($\circ$)} & \colhead{($\circ$)} & \colhead{($10^{20}\rm\;cm^{-2}$)} & \colhead{($\circ$)} & \colhead{(kpc)} & \colhead{(kpc)} & \colhead{($10^5$ M$_{\odot}$)} & \colhead{(3$'$)} &  \colhead{(3$'$)}& \colhead{(7$'$)} & \colhead{(7$'$)}
}
\startdata
1  :HVC 113.0-63.2-310\tablenotemark{e}  &  8.40 & -0.67 & 0.03 & 0.30 &  50 &  0.26 & 0.02 & 28.0 &  0.6 & 28.9 &  0.3 \nl\smallskip
& & & & & 300 &  1.57 & 0.57 & 29.5  & 2.3 & 30.4 &  1.1 \nl
2  HVC 124.7-61.9-28  & 13.70 &  0.92 & 0.02  &  0.40 & 50 &  0.35 & 0.01 & 28.0 &  0.4 & 29.0 &  0.2 \nl\smallskip
 & & & & & 300 &  2.09 & 0.38 &29.5 &  1.6&  30.5 &  0.7 \nl
3  HVC 158.1-58.6-298 &  30.33 &  -0.57 & 0.04 & 0.20 &  50 &  0.17 &  0.03 & 28.1 &  1.2  & 29.0  &  0.5 \nl\smallskip
& & & & & 300 & 1.05 & 1.08 &29.6  &  5.0  & 30.6 &  2.3  \nl
4  HVC 160.6-57.5-296 &  32.03 & -0.38 &0.06 & 0.30 & 50 &  0.26 & 0.12 & 28.1   & 5.1  & 29.0  &  2.2 \nl\smallskip
& & & & &  300  & 1.57 & 4.46 &29.6  & 20.8  & 30.6  &  9.6  \nl
5  HVC 167.0-55.6-172 &   36.00 & -0.85 &0.02 & 0.20 & 50 & 0.17 & 0.01 & 28.1   & 0.4  & 29.0   & 0.2 \nl\smallskip
& & & & &  300 & 1.05 & 0.38 &29.6  &  1.8  & 30.6  &  0.8  \nl
6  HVC 170.7-52.2-193 &  39.88 &  0.38 &0.36 &  0.20 & 50 &  0.17 & 0.36 & 28.1 &  15.1  & 29.1  &  6.7 \nl\smallskip
& & & & &  300 & 1.05 & 13.08 & 29.6  & 63.0  & 30.6  & 28.9 \nl
7  :HVC 171.0-53.6-237\tablenotemark{e} &  39.05 & -0.82 &0.41 & 0.50 & 50 &  0.44 & 1.19 & 28.1 & 49.3 & 29.1 & 21.8  \nl\smallskip
& & & & &  300 & 2.62 & 42.82 & 29.6 & 205.4 & 30.6 & 94.4 \nl
8  CHVC 173.3-52.1-225 &  41.12 & -0.60 &0.22 & 0.30 & 50 & 0.26 & 0.25 &  28.1 & 10.4 & 29.1 &  4.6 \nl\smallskip
& & & & &  300 & 1.57 & 9.01 & 29.7 & 43.5 & 30.6 & 20.0  \nl
9  HVC 254.1+53.3+105 & 166.15 &  0.88 &0.25 & 0.30 & 50 & 0.26 & 1.18 &  28.0 & 43.0 & 28.8 & 18.0 \nl\smallskip
& & & & &  300 & 1.57 & 42.48 &  29.4 & 160.0 & 30.4 & 72.7 \nl
10  HVC 261+55+124  &  170.60 & -0.08 &0.23 & 3.50 & 50 &  3.05 & 1.85 & 27.9  & 65.9 & 28.8 & 27.4 \nl\smallskip
& & & & &  300 & 18.33 & 66.59 & 29.4 & 244.6 & 30.3 & 110.6 \nl
11 HVC 339.5+56.6-110 & 212.00 & -0.68 &0.06 &0.30 & 50  & 0.26 & 0.04 & 27.5  & 0.9 & 28.3 &  0.4 \nl\smallskip
& & & & &  300 &  1.57 & 1.44 & 29.0  & 3.7 & 29.8  & 1.5  \nl
12  HVC 081.2-54.8-279 & 350.35 &  0.57 &0.09 &0.30 & 50  & 0.26 & 0.12 & 27.9 &  4.2 & 28.8 &  1.7  \nl\smallskip
& & & & &  300 &  1.57 & 4.46 &  29.3 & 15.7 & 30.3 &  7.0 \nl
13  HVC 091.7-58.3-273 & 357.08 &  0.67 &0.10 &0.20 & 50  & 0.17 & 0.10 &  27.9 &  3.4 & 28.8 &  1.4 \nl\smallskip
& & & & &  300 &  1.05 & 3.46 & 29.4 & 12.7 & 30.3  & 5.8 \nl
\enddata
\tablenotetext{a}{from Putman et al. (2002)}
\tablenotetext{b}{Semi-major axis of the HI clouds' Full Width, Half Power ellipse}
\tablenotetext{c}{fiducial distance, not actual distance}
\tablenotetext{d}{lower limit for a purely old stellar population}
\tablenotetext{e}{:HVC designation refers to clouds that could not be unambiguously classified as a CHVC (Putman et al. 2002)}
\end{deluxetable}

\ifsubmode\pagestyle{empty}\fi


\end{document}